# Digital Twin sensors in cultural heritage applications

Franco Niccolucci [1,*] and Achille Felicetti [2]

1. VAST Lab, PIN, Piazza dell'università 1, Prato, 59100, Italy; franco.niccolucci@gmail.com
2. VAST Lab, PIN, Piazza dell'università 1, Prato, 59100, Italy; achille.felicetti@gmail.com
* Correspondence: franco.niccolucci@gmail.com

**Abstract:** The paper concerns the extension of the Heritage Digital Twin Ontology introduced in previous work to describe the reactivity of digital twins used for cultural heritage documentation by including the semantic description of sensors and activators and all the process of interacting with the real world. After analysing previous work on the use of digital twins in cultural heritage, a summary description of the Heritage Digital Twin Ontology is provided, and the existing applications of digital twins to cultural heritage are overviewed, with references to reviews summarizing the large production of scientific contributions on the topic. Then a novel ontology, named Reactive Digital Twin Ontology is described, in which sensors, activators and the decision processes are also semantically described, turning the previous synchronic approach to cultural heritage documentation into a diachronic one. Some case studies exemplify this theory.

**Keywords:** Digital Twins, cultural heritage documentation, ontologies

## 1. Introduction

Digital twins are increasingly used in all research and application fields.

As discussed in recent reviews on the use of this term in different domains [1], this concept captures the complexity of digital equivalents of the real world both in industrial and research applications. It is sometimes present as a generic term to designate a digitized model of real artefacts or, more specifically, a structured information system about reality. Also regulatory authorities are using this terminology. Notable examples are offered by the European Commission, which has introduced the idea of the Digital Twin of the Earth [2] to indicate a global environmental model, but is also using this notion rather loosely in its research programmes, calling for research projects that deal with "digital twins" in different domains to simulate research activities and carry out them in a digital framework.

The idea of introducing digital twins in land planning and built environment activities dates back to the Cambridge National Digital Twin project, which stated general principles for their application in this domain, the Gemini Principles [3].

As concerns cultural heritage, the term digital twin made its appearance in reference to digitized 3D models of heritage artefacts as a synonym of digital replica, suggesting an original literary rather than technical use of the term. The first applications of 3D visualization in cultural heritage date back to the last decade of the 20th century especially in archaeology, and were motivated by the need of illustrating the supposed pristine appearance of remains, replacing drawings and maquettes, previously used for this purpose, with visual reconstructions of the past. A similar approach rapidly extended to document the shape of artefacts, monuments and sites and has by now produced a huge amount of heritage 3D models of extremely different quality and detail. Looking at heritage assets in 3D rather than by two-dimensional images has become a straightforward practice, although still needing more work as demonstrated by the EU Recommendation on this regard [4] calling for activities incentivizing the creation of new ones.

In time, it has become apparent that taking into account only the shape of heritage assets is not sufficient for an advanced model use, both in research and in practice. Any serious use requires background documentation which needs to be stored, linked to the 3D model and made available for inspection.

Work on this regard started within the 3D-COFORM project (2008-2012) [5, 6 ,7] and continued with various significant contributions especially for heritage conservation and restoration [8, 9]. In some of such applications the additional information concerning, for example, materials used in different parts or conservation interventions has been attached as a text to the 3D model or to some of its regions. Although availing of a completely different 3D technology, similar applications have been developed adopting a CAD approach to the creation of the 3D model. In this case the heritage related documentation is managed by extending an existing information system, called BIM (Built Information Modeling), widely used in the building industry to specify information about materials, services and processes concerning the construction of a *new* built asset. BIM is semantically defined by the IFC (Industry foundation Classes), which have been extended in various way to incorporate concepts typical of cultural heritage. The resulting application, called HBIM where H stands for Heritage, has a widespread use especially to document monuments and historic buildings. A 2023 paper [10] surveys a very large number of such applications. Within HBIM models, digital twins have been envisaged as a way of introducing heritage-related concepts [11, 12]. However, the data management software managing HBIM suffers from the limits of the data management software available for BIM, which privileges the analysis of individual assets modelled but is much less efficient in investigations across different buildings modelled in this way, for example those built with the same material (wood, bricks, etc.).

Both such approaches suffer from their "original sin", i.e. having a 3D model as the root of the documentation tree, where any other information is just a leave. Thus, information structures pertain to an individual heritage asset, what makes further data processing a cumbersome task: comparisons among different assets, e.g. to search for construction materials across non-visual documentation, is impractical, as no advanced database technology can assist in this job because the data are not properly organized. On the contrary, well-structured heritage documentation organizes data in *classes* of which the information pertaining to an individual asset is an instance. Classes are interrelated by *properties*. For example, "wood" would be an instance of the class "Construction Material", linked to the class "Heritage Asset" by a property "is made of". This enables searching for all assets having wood as construction material and listing the corresponding instances, i.e. all the buildings made of wood.

Classes and properties are organized in ontologies, and for cultural heritage the standard one is the CIDOC Conceptual Reference Model (CRM), ISO 21127 [13]. CIDOC stands for the committee which originally started the CRM definition, now managed by a Special Interest Group. Besides organizing the information related to cultural heritage, the CRM presents an important feature, the *extensibility*, i.e. the possibility of specializing concepts, i.e. the classes, and relationships, i.e. the properties, to fit better to the data description in a particular application subdomain. The resulting models are interoperable with any other one compatible with the CRM by establishing a correspondence (a *mapping*) between couples of corresponding classes (and properties) in each model, either directly to each other when feasible, or via the CRM class/property which is a common superclass/superproperty of both. Thus the CRM base – the universal set of heritage-related concepts – has developed several extensions, among others CRMarchaeo for archaeological excavation, CRMsci for scientific observations, CRMdig for digital activities and components, and so on, offering a complete framework to organize the knowledge about heritage. It also encompasses concepts related to intangible heritage by means of classes already proposed for narration. Both classes and properties are identified by a literal mark common to all in the ontology, followed by a progressive number; for example E followed by a number (e.g. E1) denotes the CRM classes and P followed by a number (e.g. P9) is used for properties.

The above-mentioned critical points of 3D based systems are among the reasons why in 2022 and 2023 we proposed a novel ontology for heritage information as a CRM extension. It is based on the Heritage Digital Twin (HDT), a holistic approach to heritage information where the 3D graphical component is just one element [14, 15]. In this ontology, named HDTO, all the documentation is organized according to a compatible extension of the basic CRM model and thus information from other documentation systems organized according to the CRM can be straightforwardly imported into it. Moreover, the HDTO enables documenting the intangible component of tangible heritage assets, and in the particular case of intangible heritage, such as the one enlisted in the UNESCO list [16], it is the only existing one providing a complete coverage: for intangible heritage, the visual documentation may consist in video or audio recordings or even be totally absent. 3D models have also been used to document aspects of intangible heritage (see [17] for a review of such applications), but they are usually less suitable as the starting point of the related documentation which includes stories, traditions and other immaterial content very often unrelated to the shape.

The HDTO has been used to set up the cloud-based Knowledge Base (KB) created in 4CH [18], an EU-funded project designing a Competence Centre for the Conservation of Cultural Heritage. Documentation in the 4CH KB includes the relevant information about heritage assets, from the visual and 3D one to the results of scientific analyses, conservation activities and historical research.

However, the HDTO does not (yet) consider the dynamic and interactive aspects connecting a digital twin to reality: timewise, it is static, just incorporating in each documentation item a timestamp that may be used to reconstruct the diachronic evolution of the asset. In both our previous papers, mentioned above, the authors acknowledged that this approach was just a first step in the definition of a full-fledged digital twin, still lacking the modelling of interactions with reality-based information.

In the present paper, and in other forthcoming related ones, we will try to move beyond this limitation and to develop an extended documentation system where interaction with reality may be documented and put to work in a continuous interchange of information and process activation between the real world and its equivalent in the digital universe. Such extension is motivated by the need to reflect one of the features embedded in the digital twin concept, its reactivity to inputs coming from reality and its capacity of producing corresponding outputs and real-world actions. Besides this theoretical necessity, we believe that a reactive heritage digital twin may better model heritage assets which are immersed in a continuous changing landscape, are affected by phenomena happening in it and with their own changes contribute to compose its evolution. The new semantic model proposed here extends but in no way supersedes the previous version, which conserves its usefulness in the many cases in which this dynamic perspective is not required. The HDTO is anyway the substrate on which the digitally simulated reactions take place, and the new ontology named Reactive Heritage Digital Twin Ontology (RHDTO) is an extension incorporating all the previously defined classes and properties.

Due to the complexity of such model extension, we will proceed by steps with separate contributions, the present one being dedicated to *sensors* and *activators*. This stepwise approach will also allow more focused discussion within the research community, hopefully producing improvements to be incorporated in future versions of the ontology. We will often use very simple examples to keep the reader's attention on the ontological modelling rather than on the technical aspects of the sensor, and to facilitate the model explanation to non-technical heritage professionals who play a very important role in use-inspired [19] research in this domain: but of course the ontology works also in more complicate cases as it will be illustrated in the sequel.

**2. A summary of the HDTO ontology**

To keep this paper self-contained, we summarize below the main features of the HDTO ontology. For full details, please refer to the above-mentioned papers [14, 15], introducing the concept of the Heritage Digital Twin (HDT).

According to [20] "a digital twin is a virtual representation of a physical system (and its associated environment and processes) that is updated through the exchange of information between the physical and virtual system". Another definition [21] describes the Digital Twin as made across five dimensions: Physical Entities, Virtual Models, Services, Data, and Connections.

Both our papers [14, 15] acknowledged that what they proposed was only half-way in the work needed to develop a full-fledged concept of digital twins related to cultural heritage, as they addressed only three of the above-mentioned aspects of the digital twin: Physical Entities (renamed as Real-world Entities to encompass also intangible aspects), Virtual Models, and Data, while Services and Connections were temporarily set apart to be considered in future work, of which the present paper is an initial part. This led to a formal semantic definition of the HDT and of the related ontology HDTO. The main reason for this partial move was to start setting up an overarching data infrastructure where all the parts of the Virtual Model include the relationships to a well-organized data system. The HDTO ontology is thus conceived as an extension of the CRM, enabling semantic interoperability with all the (many) existing information systems for cultural heritage documentation which adopt this standard or compatible systems, among others the EDM Europeana Data Model [22] and its forthcoming extension expected to be used in the Data Space for Cultural Heritage. Our above-mentioned papers were well received by the research community, having been extensively quoted in the literature and the first one [14] being the recipient of the *Data* 2022 Best Paper Award.

The most important innovation proposed in the HDTO is to define the concept of Heritage Digital Twin (HDT) in a holistic perspective, i.e. as *the complex of all available digital information concerning a (real world) heritage asset*, either movable (e.g. museum exhibits), immovable (e.g. monuments and sites) or even intangible (e.g. traditions). In the resulting data organization, all the data such as reports, documents, datasets and visual representations (2D, 3D or 4D), or any other related digital data, are linked with each other within the HDT by appropriate properties. Thus, the HDT becomes a full-fledged digital alter ego of real-world heritage assets.

As presented in the two papers [14, 15] mentioned above, the HDTO aims to encapsulate the dual aspects of cultural heritage, both tangible and intangible, and the cultural events in which cultural entities are involved along their history, offering specific classes and properties for the dynamic documentation and analysis of their interactions. Moreover, the HDTO is pivotal in structurally organising data to document digital twin systems, knowledge bases, and other similar operational platforms. Being machine-readable and actionable, the HDTO functions as an internal "language" of the digital twin system, ensuring smooth interaction among its various digital components. The HDTO classes describe the more common high-profile entities, such as places, agents, physical objects, events, and temporal entities, thereby ensuring high-level interoperability in multiple domains. New classes and properties are defined to describe more specific concepts such as digital twins, cultural entities, 3D models, and others, having no exact match in the CRM ecosystem. To facilitate distinguishing classes and properties in the text, henceforth we will indicate their names and symbols in ***italic boldface***.

The root class of HDTO is ***HC1 Heritage Entity***, comprising tangible and intangible entities of the real world regarded as valuable because of their contribution to society, knowledge and/or culture. ***HC1*** is a subclass of ***E1 CRM Entity***, the root class of the CRM. Instances of ***HC1 Heritage Entity*** may refer to real assets of any nature: physical, both movable and immovable, immaterial, or born digital. Subclasses of ***HC1 Heritage Entity*** model the material and immaterial aspects a cultural entity can have. In particular, the ***HC3 Tangible Aspect*** class models the tangible, material component of entities of the real-world, both movable (e.g. archaeological, artistic, and cultural objects) and immovable (e.g., built heritage as monuments, buildings, cities and other complexes). *HC4 Intangible Aspect* comprises instead cultural events, traditions and practices having particular social, historical and cultural significance, including practices and expressions, memories and oral traditions about events, things, people. They can be independent or related to some physical object. Other classes in the HDTO model the digital documentation related to heritage entities, encompassing 3D models (rendered through the dedicated *HC8 3D Model* class), images, and other audio-visual mediums (modelled using the *HC7 Digital Visual Object* class), as well as textual information made available in digital format (represented by means of the *HC6 Digital Heritage Document* class). Stories and storytelling, intended respectively as accounts of facts about cultural entities and the way in which these facts are narrated, are modelled using the narrative and narration classes provided by the Narrative Ontology (NOnt), another CRM compatible ontology for to the description of texts and their content [23]. The use of the NOnt ontology, as well as other compatible models such as CRMsci and CRMdig, demonstrates the capability of the HDTO to integrate semantic tools defined in other contexts to represent the different facets of a Digital Twin. However, no classes or properties on NOnt are used in this paper, so we refer to [15] for definitions and examples.

Heritage Digital Twins of cultural entities are rendered in HDTO by means of the class *HC2 Digital Twin,* used to organize and connect the digital information available in a given system and pertaining to an *HC1 Heritage Entity*, including digital representations, information of the effects of events that influenced or/and are related in any way to its state and activities (e.g. restorations, conservations etc.) carried out on it. The *HC2* class semantically represents the way in which the Heritage Digital Twin is implemented and the building blocks through which it operates. In addition to those inherited from the CRM, the HDTO also provides other properties, such as the *HP1 is digital twin of,* linking a heritage entity (*HC1*) with its digital twin (*HC2*). In general, the HDTO properties are mainly intended to semantically interconnect with each other the various pieces of information modelled through the classes. Thus, they are elements to implement the knowledge graph of the Heritage Digital Twin. The main classes of the HDTO and the properties linking them are shown in the following diagram (Figure 1).

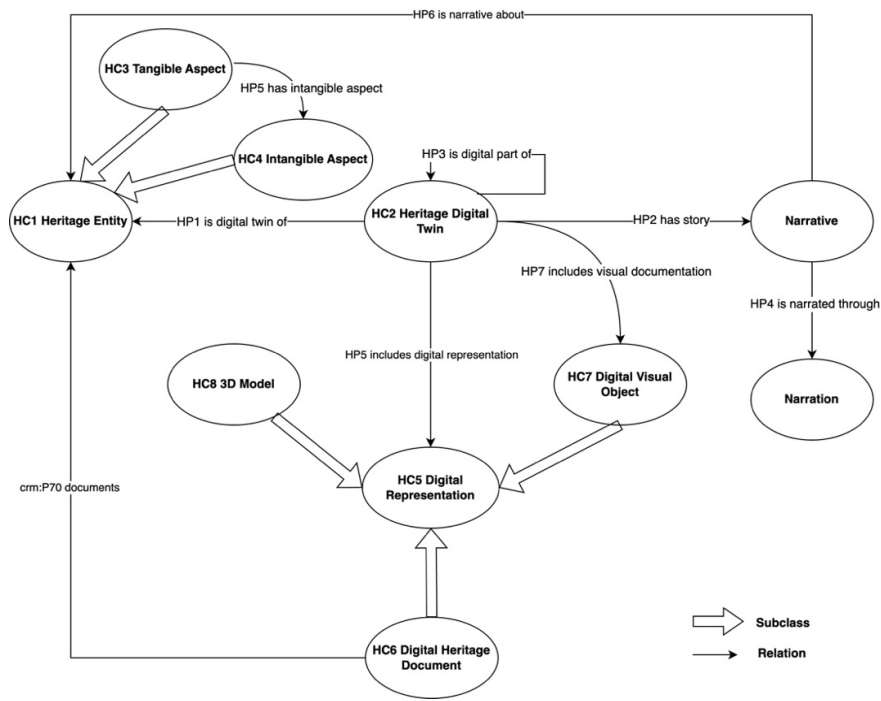

**Figure 1.** Semantic diagram of the HDTO, showing its main classes and properties.

An additional advantage of the introduction of the HDT derives from the already mentioned modelling of historical buildings using BIM (Building Information Modelling), based on the Industry Foundation Classes (IFC), and the HBIM (Heritage BIM) based on them. The extensible nature of the HDTO allows incorporating BIM and such extensions into it, via mappings from/to IFC and the proposed HBIM extensions to the HDTO.

### 3. Sensors and activators

In a simple definition, a sensor is a device that produces an output signal for the purpose of detecting a physical phenomenon. More precisely, a *sensor* is a device, module, machine, or subsystem that detects events or changes in its environment and sends the information to other electronics, frequently a computer processor. Thus, a *sensor* measures one (or more) physical quantity and sends a signal based on the measured value(s) to an *activator*, which automatically activates some action if required according to built-in criteria. Very simple everyday examples are thermostats, which combine a temperature sensor and an activator switching off or on an air conditioner if the temperature is within or outside a pre-selected range. The automated action may be very simple, as the switching on/off of a warning red light in a car fuel gauge, or an anti-theft alarm which detects the interruption of an electrical circuit due to the opening of a house front door and activates an alarm siren. The latter case may also consist in a movement detector (the sensor) that sends a signal to the siren activator. This second example is a bit more sophisticated as it may take into account other aspects, for example the volume of the intrusive object and decide if it is a thief or just your cat wandering around during its human's absence.

In any case the model of the overall action consists in several steps:

- A *sensor*: measuring one (or more) physical quantities – in the above examples they respectively are the temperature, the quantity of fuel in the car tank, and the amount of infrared radiation for an anti-theft movement sensor – and forward the measure(s) to the decider
- A *decider*: comparing the sensor measurement with pre-set decision rules and deciding if some action needs to be activated. In the above examples, respectively, for the thermostat the decision is based on a threshold, a limit temperature, and the required action is switching the air conditioning on/off; for the fuel gauge, the threshold is a minimum acceptable quantity of it, and the action is switching on the red warning light in the car dashboard; and for the anti-theft system, determining if the size of the intruder as resulting from the infrared size is comparable with a human or is smaller as the cat's one. If the sensor signal falls in the alarm range, the decider sends an activation signal to the activator. The decision rules will often be much more complex and imply different actions according to the values measured by the sensor.
- An *activator*: when authorized by the decider, it commands actions in the real world, e.g. switching some device on or off, as the air conditioning system, the car red warning light and the alarm siren. The action may be programmed to result as more complex than a simple "Do": in this case we may logically split such a complex activator into different elementary activators. For example, an anti-theft activator may command different actions such as "activate the siren" and "call police": thus it corresponds to two elementary activators.

In simple cases, the decider is included either in the sensor or in the activator, frequently in the latter as in the above examples. It is however a logical functionality different from the device, thus we will keep it separate from both.

The logic of the system sensor-decider-activator is often hard-coded in the electronics of the devices, especially for simple ones, but there is a strong tendency towards softcoding, also to support the integration with other smart devices and enable remote control via wi-fi and internet. This has led to an increase of the purely digital component, which facilitates the inclusion of these system in a digital twin framework. In any case, both the sensor and the activator are placed in the real world, so they do not belong to the digital twin, but should be included in the overall semantic description of the whole system. The decision process may be very simple as in the above examples or very complicate when several factors must be taken into account, requiring processing and possibly, in a forthcoming future, an AI-based decision process.

## 4. The Internet of Cultural Things (IoCT)

The Internet of Cultural Things, i.e. the use of the so-called Internet-of-Things (IoT) in cultural heritage applications, has been a subject of study and research in the last 10 years.

The concept of the Internet of Cultural Things (IoCT) refers to the application of Internet of Things (IoT) technologies in the context of cultural heritage and cultural assets (see [24] for a recent survey on the topic). It involves the integration of sensors, devices, and information systems within cultural contexts to monitor, preserve, study, and enhance artworks, archaeological sites, museums, monuments, and other elements of cultural heritage. In the development of digital twins in the cultural heritage sector, IoCT plays a fundamental role. In the cultural context, digital twins have extensively been used to create a digital replica of an artwork, an archaeological site, or a museum environment, allowing cultural operators to explore, preserve, and share cultural heritage in innovative ways. Sensors range from the very simple ones previously described to a network of devices measuring physical quantities and exchanging the related information to each other. Both sensors and activators may be directly linked to the digital twin, or be connected in a network in which they are the nodes.

The orchestration of such IoCT nodes may envisage different ways in which devices and sensors within an IoCT network communicate with each other and with data processing systems, allowing devices and sensors within an IoCT network to effectively and efficiently collaborate to collect, transmit, simulate and process relevant cultural data, thereby contributing to the development of digital twins and innovation in the cultural heritage sector.

Among the many studies concerning IoCT and digital twin applications in the cultural heritage domain, [25] reviews the most important applications mainly focusing on the valorisation of cultural heritage and to assist visitors and promote tourism. Very simple examples of such applications are the sensors activating a video when somebody enters a museum room or in general to start an automated reaction to human presence, to make the heritage enjoyment more interactive and participative for visitors. More complex ones collect measures from various devices, connected in an articulated system as described above, and use these to take decisions on actions to be performed by one or more activators.

Digital twin applications to the conservation of monuments are discussed in general in [26], while [27] describes the use of sensors to monitor the conservation conditions of historic buildings, in this case a church in Matera's "Sassi" considered in its complex environment. Applications to movable artefacts are less frequent, among them the one described in [28] to a famous violin and the study of palaeolithic lithics presented in [29]. A semantic approach to digital twins is introduced in [30], which discusses the ontologies for the description of sensors and IoT in preventive conservation. Other related papers are [31], [32], [33] and [34].

## 5. The Reactive Heritage Digital Twin Ontology (RHDTO)

### 5.1. Introductory Notes

In this section, we introduce the new entities of the RHDT Ontology aimed at modelling the reactive elements of the Digital Twin, such as sensors, services, and data processing engines. As mentioned earlier, our ontology builds upon the CRM ecosystem and incorporates elements from the CRMdig, CRMsci, and CRMpe extensions. Concepts and logics from various existing ontologies, developed to model sensors and related phenomena, have been investigated to fine tune our conceptual tools. Models such OntoSensor [31], and SensorML [32, 33] provide a rich and expressive framework for representing sensors, their properties, and their relationships to other entities, such as observed properties, features of interest, and observation processes. Nevertheless, our primary goal is not only to model the phenomena related to digital objects, their nature and their interactions, but also to develop a semantic tool for making Reactive Digital Twins operational and achieving integration and interoperability between the different types of data they store and manage.

By leveraging the CRM ontology and extending it with new classes and properties where necessary, we aim to create a comprehensive ontology that can effectively represent the dynamic and complex nature of cultural heritage and risk management, and support decision-making and risk assessment in real-world scenarios. CRM, with its robust framework for modelling events, is particularly well-suited for capturing the dynamic aspects of the Reactive Heritage Digital Twins and representing the complexity typical of this domain, including state changes detected by sensors and other devices, component interactions and processes. Leveraging this event-centred approach allowed us to better focus on system behaviours and predict future outcomes by identifying common patterns and trends within the system's operations.

Throughout the development process, we have thus prioritized the reuse of existing classes and properties from the CRM ecosystem, only introducing new ones when necessary to accurately represent the intended semantics of the described entities. This approach strikes a balance between ensuring interoperability and maintaining precision in describing the functionalities and components of the system, while also keeping the ontology compact and easy to understand and apply. Therefore, the resulting RHDT ontology only introduces classes and properties to describe the specific components and events involved in the various operations performed by the system, the interaction between devices and services, and the communication with human operators.

The following Table 1 reports the symbols used for classes and property names in the CRM and its extensions used in the present paper. For example, in the CRM the class symbol is E and the property symbols is P. Symbols are combined with a number to identify a specific class or property, e.g. *E1* or *P2*.

**Table 1.** CRM extensions used in the paper.

| Extension name | Used for | Reference | Class symbol | Property symbol |
|---|---|---|---|---|
| CRM | General use | [13] | *E* | *P* |
| CRMdig | Digital objects | [35] | *D* | *L* |
| CRMsci | Scientific analyses | [36] | *S* | *O* |
| CRMpe | Interoperability framework | [37] | *PE* | *PP* |
| HDTO | Heritage digital twins | [14, 15] | *HC* | *HP* |

Also in the following sections we will use the convention of denoting the semantic elements in the text by *italic-boldface*.

Another extension of the CRM, proposed in [38], describes a model for the results of scientific analyses on heritage assets to be used for the data concerning them. It is not used in the present paper, but its classes and properties are relevant in applications to support the description of condition states of artefacts, of the experiments carried out to evaluate them and their results.

*5.2 Sensors*

Sensors are a central element of the Reactive Digital Twin. In our ontological view, a sensor can be defined in general terms as a digital device placed on physical objects or in specific locations, intended to measure and collect data about them, processes it, and transmit it to the Digital Twin system for analysis and further processing. A sensor can measure various physical or environmental properties, such as temperature, pressure, humidity, light, sound, position, velocity, or acceleration, and can be of different types, such as analogic or digital, wired or wireless, active or passive. Building upon this general definition, we introduced the new *HC9 Sensor* class, specializing the *D8 Digital Device* of CRMdig, a general class aimed at describing instances of material items capable of processing or producing digital data.

The positioning of a sensor on a cultural object can be modelled using the new *HP15 is positioned on* property, allowing for the representation of the spatial relationship established between the object to be monitored (*HC3 Tangible Aspect*) and the sensor placed on it. For sensors positioned in a space adjacent to the cultural object, we can use the *P55 has current location* property of the CRM, connecting a sensor to the instance of *E53 Place* where the monitored object is currently located, to indicate that both these physical entities share the same space.

During its operational lifecycle, a sensor can assume a particular status representing its condition at a given time, which can be modelled using the *E3 Condition State* class of CRM. Instances of this class enable the description of sensor states (such as on/off, operational/non-operational) and operation modes, reporting any faults or errors, as well as any maintenance or calibration activities performed on it. Additionally, sensors are typically operated by different kinds of software, tailored for gathering and processing acquired data, generating signals, and executing specific functions such as configuration, calibration, monitoring, and diagnostics. To represent this operational software, we employ the *D14 Software* class of CRMdig, that was chosen due to its comprehensive semantic features allowing for the modelling of all the software components of the Reactive Digital Twin. While the *D14* class already provides the necessary features to model the software components of any device, we deemed important to explicitly represent the relationship between the sensor and its controlling software. For this reason, a specific *HP11 is operated by* property was created to indicate the close interconnection existing between the sensor and the software that controls it. This also enables the link between the software, the measurement operations performed by the sensor and the digital signals generated out of them.

*5.3 Measurements and Signals*

In a CRM perspective, the measurements operations performed by sensors can be modeled as events. A sensor measurement event, in fact, involves the collection of data identifying various modifications of conditions on the object or in the environment under examination, within a certain spatiotemporal interval. In accordance with these features, we have defined the new *HC13 Sensor Measurement* class as a subclass of the *S21 Measurement* class of the CRMsci ontology, to render the measurement events performed by the sensors connected to Reactive Digital Twins. The *L12 happened on device* property of CRMsci is used to specify the sensor (*HC9*) on which the measurement (*HC13*) took place. Measured events can instead be modelled by means of the CRM *E5 Event*, a class particularly well-suited for describing generic events of various types. The measurement (*HC13*) and the measured (*E5*) events are ontologically related through the *O24 measured* property of CRMsci. Instances of the CRM *E55 Type* class can be assigned to instances of *HC13* to detail, among the other things, the detected types of risk conditions related to the cultural object, and subsequently trigger the generation of the corresponding signals. This assignment is usually done by means of the *L17 measured thing of type* property of CRMsci.

Generated signals can be modelled as digital objects since they codify a measurement taken by a sensor to be transmitted, under certain conditions, to the system for further processing or analysis. Typically, a signal is a piece of software encoded in a formal language (e.g., XML, JavaScript, etc.) and generated in response to a specific event, such as a change in sensor reading, a threshold crossing, or a specific timer. It may contain various information, including sensor identifiers, timestamps, recorded values, measurement units, and details about condition assessment quality. In our model, instances of signals are represented by means of the new *HC12 Signal* class, a specialization of the *D9 Data Object* class of the CRMdig ontology. Signals (*H12*) generated in the presence of a potential risk detected by sensors can be linked to the measurement event that generated them through the *L20 has created* property of CRMdig. Once generated, signals are transmitted to the Reactive Digital Twin, and specifically to dedicated intelligent digital agents (deciders) running as part of the Reactive Digital Twin system. The transmission of signals to deciders is recorded via the new *HP12 was transmitted to* property.

*5.4 Deciders*

RHDTO provides a new *HC10 Decider* class, specializing the scope of *PE1 Service* of the CRMpe ontology, to describe in detail the deciders and their features. Deciders can receive and analyse input data from various sources, such as sensors, activators, or other services, and query the Digital Twin Knowledge Base to acquire knowledge concerning the linked cultural objects and their potential risks. They further process all this information using algorithms, rules, or models, and generate output instructions, such as commands, feedback, or status updates, to be sent to other components of the Digital Twin, to activators or other external devices. Moreover, deciders can also send alerts to human operators in various forms, such as email, SMS, push notifications, visual or audible alerts, depending on the nature of the detected issue.

Ontologically, the actions performed by deciders after the decision-making process is complete can be modelled as events. Thus, we have designed a specific *HC14 Activation Event* class, subclass of the CRM *E5 Event*, to represent them. The *O13 triggered* property of CRMsci is particularly useful for linking activation events to the deciders by which they were triggered. Activation events can prompt specific actions by digital agents, such as adjusting a valve or activating a pump, and/or alert human operators, informing them of the necessary countermeasures to be taken in response to the detected potential risks. Both operations are modelled by means of the new properties *HP14 alerted*, pointing to instances of the *E39 Actor* class, and the *HP13 activated*, linking activation events with activators.

*5.5 Activators and human interaction*

Activators are other fundamental components of Reactive Digital Twins. In our RHDT ontology they are described as devices that enact actions upon physical objects or processes, based on the instructions received by the deciders operating within the Reactive Digital Twin system. They can be of mechanical, electrical, digital, hydraulic, or pneumatic types, and can execute a range of actions, including movement, adjustment, control, or regulation. Activators may operate in distinct modes, including automatic, manual, or semi-automatic, and can be governed by various software systems, such as firmware, drivers, or applications. Moreover, they offer various degrees of precision, accuracy, and responsiveness, contingent upon their design, fabrication, and maintenance processes.

Activators interact with the Reactive Digital Twin through multiple channels, including sensors, APIs, or communication protocols, enabling the reception of commands, feedback, or status updates from the system. Given their specialized nature, we have defined a new *HC11 Activator* class, subclass of the *D8 Digital Device* of the CRMdig ontology to describe these peculiar devices. As in the case of sensors, activators may be operated by a software (again encoded through the *D14 Software* class of CRMdig) intended for interpreting the commands and feedback coming from the Reactive Digital Twin system and control the activator's actions accordingly.

As already anticipated, the *E39 Actor* CRM class is used to represent people and/or institutions responsible for safety and the security of cultural entities and their environment. These are the people who (may) receive alerts from deciders in order to take appropriate actions in case of risky events. As we said, the **HP14 alerted** property is used to associates an instance of **HC14 Activation Event** with an instance of **E39 Actor**, indicating the alerts sent to the specified individuals or groups. This property is of particular interest since it describes the modalities of interaction and collaboration between the digital system and the real world, making particularly evident, also on the ontological level, the importance of the human component for the effective prevention and management of dangerous situations.

A schematic overview of the RHDT Ontology, its entities and their semantic relationships is illustrated in Figure 2.

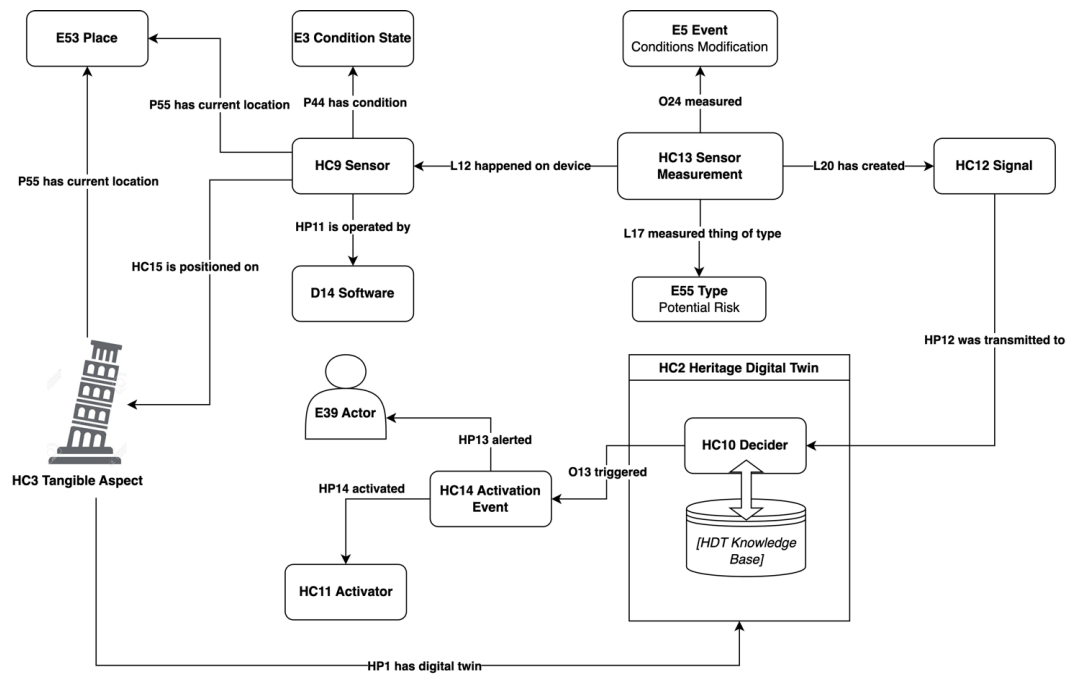

**Figure 2.** Classes and properties of the RHDT Ontology.

## 6. Summary of the new classes and properties

The new classes and properties are listed below with their scope note, i.e. a short description.

*6.1 Classes*

*HC9 Sensor*
Subclass of: **D8 Digital Device** of the CRMdig ontology.

This class comprises specialized devices utilized for monitoring and evaluating the conditions of artifacts, structures, or environments of historical significance. These devices are designed to detect changes in parameters such as temperature, humidity, vibration, or light exposure, which may impact the conservation of cultural assets. Operated by dedicated software, sensors can exist in various states, including active, standby, or alert. For example, a temperature sensor may indicate fluctuations that could endanger delicate artifacts like manuscripts or artworks. When certain predefined thresholds are surpassed, the sensor enters an alert state, signaling potential risks and prompting appropriate intervention measures.

*HC10 Decider*
Subclass of: **PE1 Service** of the CRMpe ontology.

This class comprises software components responsible for receiving signals from sensors and autonomously making decisions aimed at safeguarding cultural objects. Acting as an intelligent service within the conservation systems of the Digital Twin, deciders process incoming data from sensors and information stored in the Digital Twin Knowledge Base to detect factors such as environmental conditions, artifact vulnerability, and conservation protocols. Based on this analysis, it executes predefined algorithms or decision-making rules to determine appropriate actions for preserving cultural heritage. These actions could include adjusting environmental controls, activating protective measures, or triggering alerts to conservators or relevant personnel when necessary.

*HC11 Activator*
Subclass of: ***D8 Digital Device*** of the CRMdig ontology

This class comprises digital devices responsible for executing actions determined by the decider of the Digital Twin system to safeguard cultural objects. Serving as a crucial link between decision-making and practical implementation, the activator translates directives from the decider into tangible interventions aimed at mitigating risks or optimizing conservation conditions. Interventions may encompass a variety of mechanisms, including the activation of mechanical, electronic, pneumatic, or hydraulic systems, among others. Activators ensure the timely and effective execution of proactive conservation strategies, contributing to the long-term preservation of cultural heritage assets.

*HC12 Signal*
Subclass of: ***D9 Data Object*** of the CRMdig ontology.

This class is used to model particular data objects generated by sensors and documenting specific detected conditions. Signals are typically transmitted to the digital twin, where they are processed and analyzed by the system's algorithms to generate insights into the condition and conservation needs of the cultural heritage. Signals may be encoded in a specific data formatting language, facilitating efficient transmission, storage, and analysis. Utilizing standardized formats ensures interoperability and compatibility among different sensor systems and conservation platforms, including the Digital Twin Knowledge Base, enabling seamless integration of data from diverse sources for comprehensive conservation management. The incorporation and analysis of encoded signals enables the Digital twin to support informed decision-making and proactive preservation strategies for cultural heritage assets.

*HC13 Sensor Measurement*
Subclass of: ***S21 Measurement*** of the CRMsci ontology.

This class comprises specific measurement events in which a sensor detects and quantifies a specific parameter or condition relevant to the monitoring and conservation of cultural assets. Events of this kind occur when a sensor registers changes in parameters such as temperature, humidity, light exposure, or vibration, capturing information that reflect the environmental conditions surrounding cultural objects or structures. The sensor measurement is thus an essential event for continuously assessing the time and circumstances in which conditions of risk may affect cultural heritage, providing valuable insights into factors that may impact the preservation of artifacts or sites over time.

*HC14 Activation Event*
Subclass of: ***E5 Event*** of the CRM ontology.

This class serves to model actions performed by an activator to initiate specific interventions or alerts aimed at safeguarding cultural assets. Activation events occur when an activator executes directives received from the Digital Twin's decider, triggering actions such as activating climate control systems, deploying protective enclosures, or alerting personnel through various communication channels, including email notifications, SMS and similar.

*6.2 Properties*

**HP11 is operated by**
Domain: ***HC9 Sensor***

Range: *D14 Software*

This property links an instance of *HC9 Sensor* the instances of *D14 Software* that operate it. Software is usually a piece of code running on the sensor and responsible for controlling and managing it, for instance by configuring its settings, collecting and processing its data, and generating signals based on its measurements.

*HP12 was transmitted to*:
Domain: *HC11 Signal*
Range: *HC9 Decider*

This property associates instances of *HC11 Signal* with instances of *HC9 Decider*, indicating that a certain signal has been transmitted to the Digital Twin decider for processing. The property can be used to model the flow of data from sensors to the deciders and can be useful for tracking the status of signals and ensuring that they are properly processed by the decider. The property can also be used to model the relationship between signals and the specific decider services that process them, allowing for more fine-grained analysis and optimization of the Digital Twin system.

*HP13 activated:*
Domain: *HC14 Activation Event*
Range: *HC11 Activator*

This property associates instances of *HC14 Activation Event* with instances of *HC11 Activator*, indicating the specific digital device activated by the Digital Twin system based on the decision made by the decider component regarding the actions to be taken according to the detected risk.

*HP14 alerted:*
Domain: *HC14 Activation Event*
Range: *E39 Actor*

This property associates instances of *HC14 Activation Event* with instances of *E39 Actor*, indicating the action of alerting human personnel. The property can be used to model the communication between the Digital Twin and human operators, following the decision taken by the decider component, and to describe the modalities of collaboration and decision-making between the digital system and the real world.

*HP15 is positioned on:*
Domain: *HC9 Sensor*
Range: *HC3 Tangible Aspect*

This property is used to model the spatial relationship between instances of HC9 Sensor and the physical object on which sensors are located. The property is fundamental to document the specific case in which a sensor is physically placed on or attached to a cultural heritage physical object rather than simply placed nearby or in the same environment as the monitored object.

## 7. Example: Giovanni Pisano's pulpit in Pistoia, Italy

To exemplify the documentation of a system using sensors, we will consider a recent paper concerning the pulpit in the church of Sant'Andrea in Pistoia (Italy), a medieval masterwork by the Italian sculptor Giovanni Pisano described in [39]. The physical system includes sophisticated sensors and takes into account previous work by a world-class restoration centre based in Florence, the *Opificio delle Pietre Dure* (OPD), as described in the above-mentioned paper. The array of sensors installed for the purpose of gathering environmental and dynamic data encompassed various types, among which were humidity sensors situated on the wall in the side nave adjacent to the pulpit, and uniaxial accelerometers positioned atop the pulpit and on the ground near its base. This specific configuration facilitated meticulous monitoring of both the environmental area surrounding the pulpit and any dynamic oscillations occurring on its surface.

To demonstrate the functionality of our ontology in representing this scenario, we specifically focus on these two types of sensors, illustrating how they can be semantically described using our classes and properties, and integrated into the overall description of the installed monitoring system.

The semantic modelling starts by noting that Giovanni Pisano's pulpit is a monument and thus a physical cultural object that can be represented by instantiating the *HC3 Tangible Aspect* class of our RHDT Ontology. The church of Sant'Andrea in Pistoia (Italy) can be represented by an instance of the *E53 Place* class of the CRM, while the fact that the pulpit is housed in this church can be rendered through the *P55 has current location* property of the same model. Specific identifiers for both these heritage entities can be defined and/or derived using for example the global identifiers provided by Wikidata, such as the https://www.wikidata.org/wiki/Q3925522 URI identifying the pulpit, or the https://www.wikidata.org/wiki/Q1148335 URI identifying Sant'Andrea's church in the Wikidata system. The *HC2 Digital Twin* class can be instantiated to define the Reactive Digital Twin of the pulpit in the semantic space of our model.

As previously noted, instances of the *HC9 Sensor* class could be employed to represent the diverse types of the sensors used to monitor the monument. We focus our example on two instances of this class, respectively used to represent one of the temperature and humidity sensors installed on the walls of the church, and the uniaxial accelerator sensor positioned on top of the pulpit. Regarding their specific position, the distinct placements of these two sensors are rendered either through the CRM property *P55 has current location*, used to indicate the positioning of the temperature and humidity sensor on the wall of the church, and the property *HC15 is positioned on*, used to specify the placement of the uniaxial accelerometer sensor directly on the surface of the monument. The *HP11 is operated by* property is utilized to establish the connection between the sensors and the corresponding instances of CRMdig *D14 Software* class, employed to represent the software operating them.

An associated instance of the *HC13 Sensor Measurement* class is defined to specify, through the CRMdig *O24 measured* property, the type of event monitored by each sensor, and specifically, the seismic movements detected through variations in acceleration by the uniaxial accelerometer, and changes in temperature and humidity parameters of the church recorded by the temperature and humidity sensor. Both these events are modelled by means of the *E5 Event* class of CRM.

The signals resulting from these measurements are represented by using the *HC12 Signal* class and linked to the event that generates them via the *L20 has created* property of CRMdig. The signal transmission to the Digital Twin of the pulpit (*HC2*) is encoded through the *HP12 was transmitted to* property, and the integrated monitoring system that receives them by instances of the *HC10 Decider* class. This is the system designed to acquire the transmitted values, analyse the various physical and environmental conditions of the monument, extend the information coming from other sensors, enrich and interact with the Digital Twin's Knowledge Base.

The sensor system of Giovanni Pisano's pulpit does not include activators, but just for the sake of exemplifying how the system would be described when including an activator, we have added a hypothetic second part, not present in the actual system. In this supposed system, a decider (*HC10*) evaluates the variations of some measures from standard values and if they exceed a threshold, it sends an email to the OPD to intervene. This process can be semantically modelled by means of the *O13 triggered* CRMdig property and the *HC14 Activation Event* representing the action of email transmission by the system to the OPD competent office (*E39 Actor*).

Figure 3 represents the knowledge graph of the example.

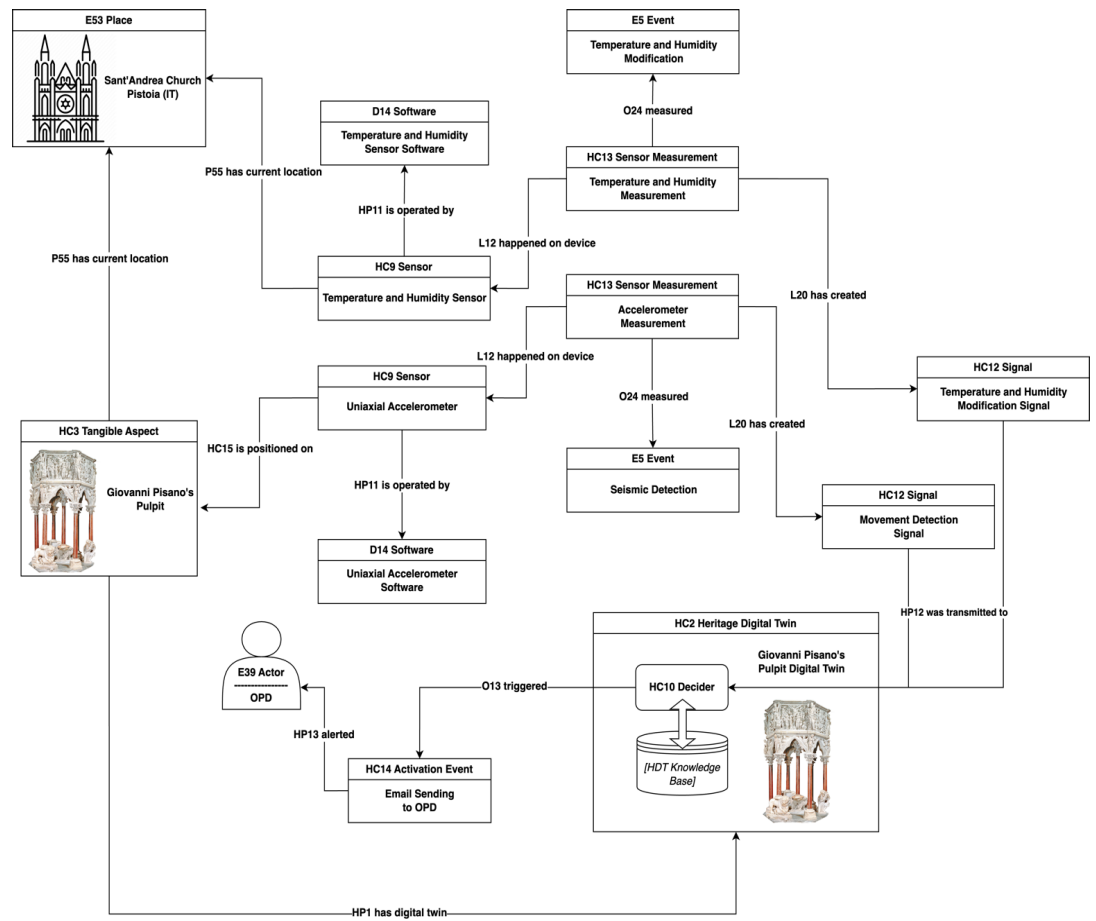

**Figure 3.** Semantic graph of the Giovanni Pisano's pulpit monitoring system using the RHDT Ontology.

## 8. Conclusions and future work

In the present paper we have addressed the introduction of sensors and activators in the RHDTO ontology, to describe the reactions of a system modelled as the digital twin of a heritage asset to external real-world events, The impact of such events is analysed in terms of measures of one or more physical quantities made by sensors, analysed by deciders which command actions (possibly no action) to activators.

This model does not cover all the possible reactions of the digital twin to external events. They may be described by information collected by external systems and forwarded to the digital twin system, such as weather forecasts or earth science information; then, combined with other data by the decider, for example with the geographic location of the asset; and ultimately lead to actions by activators in a way similar to the sensor-decider-activator chain but with the sensor replaced by the external information. A rather famous example is the MOSE system in Venice, used to limit the effects of flood, caused by high tide and strong northern winds, on the monuments of the famous city by rising mechanical barriers at the lagoon mouths on the Adriatic sea. MOSE bases its action on forecasting high tides, which have a well-known regularity, and wind strength and direction with weather forecasts. Although discussed for some anticipated environmental side effects, MOSE has so far succeeded in avoiding the effects of *acqua alta* (literally, "high water", i.e. high tide) on the city monuments.

Finally, another possibility is the simulation of external events, digitally modelled in the system and described in the digital twin, to be analysed by the decider and produce in output the simulated consequences, possibly also including the simulated effects of corrective actions by activators. For example, one might ask the system which are the effects of extremely hot summers on an art gallery and how these are mitigated by air conditioning. All such simulations will need scientific models behind the scenes and the related processing programs, which may be stored in a computer system and documented with all the other information in the corresponding Reactive Digital Twin.

**Author Contributions:** All authors have equally contributed to the present paper. They read and agreed to the pre-print version of the manuscript.

**Funding:** This research received no external funding.

**Conflict of interest:** Authors declare no conflict of interest.